\documentclass[showpacs,aps,prl,twocolumn]{revtex4-1}
\usepackage{bm} 
\usepackage{graphicx}
\usepackage{amsmath}
\usepackage{epsfig,amsmath,graphicx,amssymb,array,booktabs}
\usepackage[table]{xcolor}

\begin{document}

%\preprint{}

\title{Is hexagonal InMnO$_3$ ferroelectric?}

\author{F. -T. Huang$^1$}
\author{X. Wang$^1$}
\author{Y. S. Oh$^1$}
\author{K. Kurushima$^2$}  
\author{S. Mori$^3$}
\author{Y. Horibe$^1$} 
\author{S. -W. Cheong$^{1}$}
 \email{sangc@physics.rutgers.edu}
\affiliation{
$^1$ Rutgers Center for Emergent Materials and Department of Physics ${\&}$ Astronomy, Rutgers University, Piscataway, New Jersey 08854, USA }
\affiliation{
$^2$ Toray Research Center, Ohtsu, Shiga 520-8567, Japan}  
\affiliation{ 
$^3$ Department of Materials Science, Osaka Prefecture University, Sakai, Osaka, 599-8531, Japan}

%\date{\today}

\begin{abstract}  %600 character aorund 80 word
The presence of ferroelectricity in hexagonal (\textit{h}-)InMnO$_3$ has been highly under debate. The results of our comprehensive experiments of low-temperature
(T) polarization, TEM and HAADF-STEM on well-controlled \textit{h}-InMnO$_3$ reveal that the ground state is ferroelectric with \textit{P}6$_3$\textit{cm} symmetry,
but a non-ferroelectric \textit{P}$\bar{3}$\textit{c}1 state exists at high T, and can be quenched to room T. We also found that the ferroelectric
\textit{P}6$_3$\textit{cm} state of \textit{h}-InMnO$_3$ exhibits the domain configuration of topological vortices, as has been observed in \textit{h}-REMnO$_3$
(RE=rare earths).       
\end{abstract}
%high angle annular dark-field scanning transmission electron microscopy (

%\pacs{77.80.Dj,77.80.B-,68.37.-d}
%magnetic oxides, 75.47.Lx  
%Ferroelectricity
%domain structure, 77.80.Dj
%phase transitions, 77.80.B-  
%Electron microscopy in structure determination, 68.37.-d

\maketitle
%structural

\begin{figure}[t] 
\begin{center}
\includegraphics[width=3.3in]{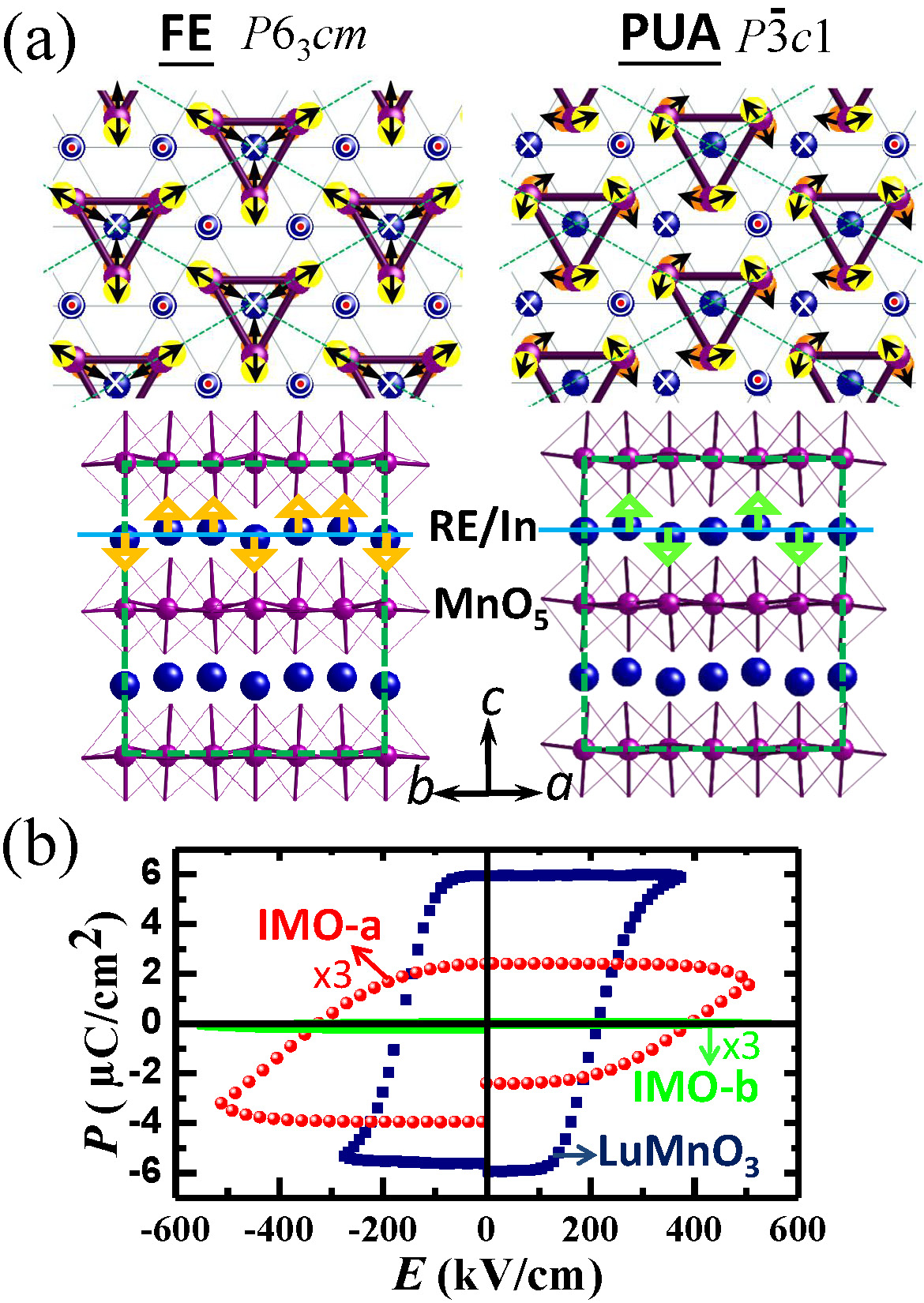} 
\end{center}
\caption{\label{fig:Fig1}(Color online) (a) The top and side view of the ferroelectric (FE, \textit{P}6$_3$\textit{cm}) and partially undisordered antiferroelectric (PUA, \textit{P}$\bar{3}$\textit{c}1) structures of \textit{h}-RE(In)MnO$_3$. The navy (large), purple (small), yellow, and orange spheres represent RE/In, Mn ions, upper, and bottom apical O ions of MnO$_5$ bipyramids, respectively. The arrows depict the directions of RE/In and O atomic distortions. The triangles with purple bars correspond to the Mn-trimers. The green dashed lines represent the unit cells of the  $\sqrt{3}$$\times$$\sqrt{3}$ superlattice due to In distortions and Mn-trimers. (b) Polarization vs electric field, \textit{P(E)}, hysteresis loops for polycrystalline IMO-\textit{a} (red spheres), IMO-\textit{b} (green solid lines) measured at 100 K, and a LuMnO$_3$ single crystal (navy squares) at 150 K.}
\end{figure}

Hexagonal manganites (\textit{h}-REMnO$_3$, RE: Y and Ho-Lu) continue to attract great attention because of its improper ferroelectricity, the presence of topological vortices and multiferroicity \cite{Choi2010, Spaldin2004, Fennie2005}. The size mismatch between RE layers and Mn-O layers induces a trimerization-type structural phase transition from the high-temperature paraelectric state (PE, \textit{P}6$_3$/\textit{mmc}). In order to achieve favorable close packing, the rigid MnO$_5$ trigonal bipyramids tilt, which leads to a loss of an inversion symmetry with 2/3 upward- and 1/3 downward-distorted (up-up-down) RE ions along the \textit{c} axisthis imbalance induces a ferroelectric state [FE, \textit{P}6$_3$\textit{cm} in Fig.~\ref{fig:Fig1}(a)] \cite{Spaldin2004}. The magnitude of the MnO$_5$-bipyramid tilting and that of the RE-layer buckling increase naturally with decreasing RE ionic radius due to increasing layer-size mismatch \cite{Katsufuji2002, Aken2004}. The various physical characteristics of such structural-driven ferroelectricity, including the phase transition temperature (Tc) and the magnitude of ferroelectric polarization are evidently coupled with the size of RE ions \cite{Katsufuji2002, Aken2004, Chae2012}. InMnO$_3$, where In ions are much smaller than any RE ions in size, does form in a similar hexagonal structure, and thus it is intriguing to find out the possible ferroelectricity in hexagonal InMnO$_3$.

Ferroelectricity in InMnO$_3$ has been highly controversial. InMnO$_3$ was theoretically predicted and experimentally claimed to show weak ferroelectricity with
Tc$\sim$500 K in 2001 \cite{Abrahams2001} and 2006 \cite{Serrao2006}, respectively. More recently, despite a fully-filled 4\textit{d} orbital in InMnO$_3$ distinct
from YMnO$_3$ \cite{Spaldin2004}, Oak \textit{et al.} proposed an alternative intra-atomic 4\textit{d}$_z^2$-5\textit{p}$_z$ orbital mixing of In and a covalent
bonding [4\textit{d}$_z^2$(In)-2\textit{p}$_z$(O)] along the \textit{c}-axis, resulting in a ferroelectric ground state \cite{Oak2011}. On the other hand, Belik
\textit{et al.} reported the absence of spontaneous polarization \cite{Belik2009}. In addition, Kumagai \textit{et al.} reported their experimental results of no
ferroelectric signals in second-harmonic generation (SHG) and piezoelectric force microscope (PFM) measurements, and concluded a non-polar structure for InMnO$_3$
\cite{FieBig2012}. Based on DFT calculations, they also claimed that the ground state is a non-ferroelectric state with the space group of
\textit{P}$\bar{3}$\textit{c}1 even though there exists only a small energy difference between the non-polar \textit{P}$\bar{3}$\textit{c}1 and ferroelectric
\textit{P}6$_3$\textit{cm} states. In the non-polar \textit{P}$\bar{3}$\textit{c}1 state, In ions sit equally on 1/3 downward, 1/3 in mirror plane and 1/3 upward (down-no-up) positions \cite{FieBig2012}. In frustrated Ising triangular (i.e. hexagonal) antiferromagnets, a PDA state refers to a partially disordered
antiferromagnetic state where spins on a honeycomb lattice portion of the hexagonal lattice order antiferromagnetically, and the rest of spins are disordered
\cite{PUA state}. Due to the evident analogy, we will call the non-polar \textit{P}$\bar{3}$\textit{c}1 as a PUA (partially un-distorted antiferroelectric) state
[Fig.~\ref{fig:Fig1}(a)]. The preparation of high-quality InMnO$_3$ and the growth of decent-size single crystals turn out to be challenging, which is partially the
origin of the controversy. Therefore, it is imperative to find out the correct crystallographic ground state in well-controlled specimens of InMnO$_3$.

%\textit{P}6$_3$\textit{cm} and \textit{P}$\bar{3}$\textit{c}1 

%since ScMnO$_3$ has been proven to be ferroelectric with an even smaller ionic radius\cite{Katsufuji2002, Abrahams2001}.  %to complete or to xxxx 
%equivocal
%illustrate two sturctures, try to compare primary cell  %can we tell from X-ray

%fig1 red arrow in sparkle+ PE loop %====================================== 
%======================================

%when anomalous scattering is present and each specimen was found to contain a small, roughly constant amount of In$_2$O$_3$
In general, x-ray and electron diffraction methods are two main powerful and robust techniques to study crystallographic structures. However, the same extinction rules
in both FE and PUA states provide difficulties to distinguish them from, for example, x-ray structural refinements, even though slightly different intensity ratios of
Bragg peaks exist due to different structural factors. On the other hand, dark-field transmission electron microscopy (DF-TEM) is an ideal and well-known technique to
examine domain structures in ferroelectric and non-ferroelectric materials because of its high spatial resolution and ability to isolate specific-type domains using a
specific diffraction spot \cite{Choi2010, Tanaka-JPSJ}. In particular, we can examine the existence of inversion symmetry by taking advantage of the Friedel's law,
where the Friedel-related pairs of Bragg reflections should behave differently in a non-centrosymmetric structure \cite{Tanaka-JPSJ, Gevers-F}. In this letter, we have
investigated the domain morphologies of a series of \textit{h}-InMnO$_3$ specimens prepared in different conditions using DF-TEM as well as high angle annular dark-field scanning transmission electron microscopy (HAADF-STEM) with atom-resolved spatial resolution. We provide conclusive
evidences of the presence of the ferroelectric ground state in InMnO$_3$ with characteristic topological vortices. In addition, we demonstrate that the long-range or
short-range FE and/or PUA states can be deliberately controlled by varying thermal treatments.

Polycrystalline specimens of InMnO$_3$ were prepared by a solid-state reaction method. The mixtures of In$_2$O$_3$ (99.999\%), MnO$_2$ (99.99\%) powders with the
stoichiometric ratio were ground together, pelletized, and then heated at 980 $^{\circ}$C for 200 hours in air. A small amount (2-5\%) of Bi$_2$O$_3$ (99.975\%)
was added to enhance the grain growth of InMnO$_3$. Four polycrystalline InMnO$_3$ specimens are discussed in this letter: IMO-\textit{a} was slowly cooled (2
$^{\circ}$C/hr) from 980 $^{\circ}$C; IMO-\textit{b} was furnace cooled; IMO-\textit{c} was quenched from 950 $^{\circ}$C; and IMO-\textit{d} was quenched from 650
$^{\circ}$C to room temperature after cooled slowly (10 $^{\circ}$C/hr) from 980 $^{\circ}$C. Polarization-electric field, \textit{P(E)}, hysteresis loops were
measured at low temperature with a programmable function generator (DS340), high voltage amplifier, and oscilloscope (TDS1010). Specimens for TEM studies were prepared
by mechanical polishing, followed by Ar ion-milling. Domain structures were studied using a JEOL-2010F transmission electron microscope equipped with a 14-bit
charge-couple-device (CCD) array detector. Imaging plates were also used to record dark-field images. All the Miller indices described in this letter are based on
\textit{P}6$_3$\textit{cm} structure. High-angle annular dark-field (HAADF) imaging and chemical mapping with an atomic-column resolution were carried out using a
JEOL-ARM200F scanning transmission electron microscope equipped with a spherical aberration Cs-corrector in conjunction with energy-dispersive x-ray
spectroscopy.

%Though, SHG signals failed to have signals from the breaking of inversion symmetry in InMnO$_3$ \cite{Katsufuji2002} and there is no anomaly feature of dielectric permittivity\cite{Belik2009}. %emphasis real space and atomic resolution and structural technique. %The main point of this paper, the technique advantage 

%reveal that the domain structures consist of irregularly shaped domains with curved domain walls.
%fig2-sparkle
%======================================
\begin{figure}[t]
\begin{center}
\includegraphics[width=2.8in]{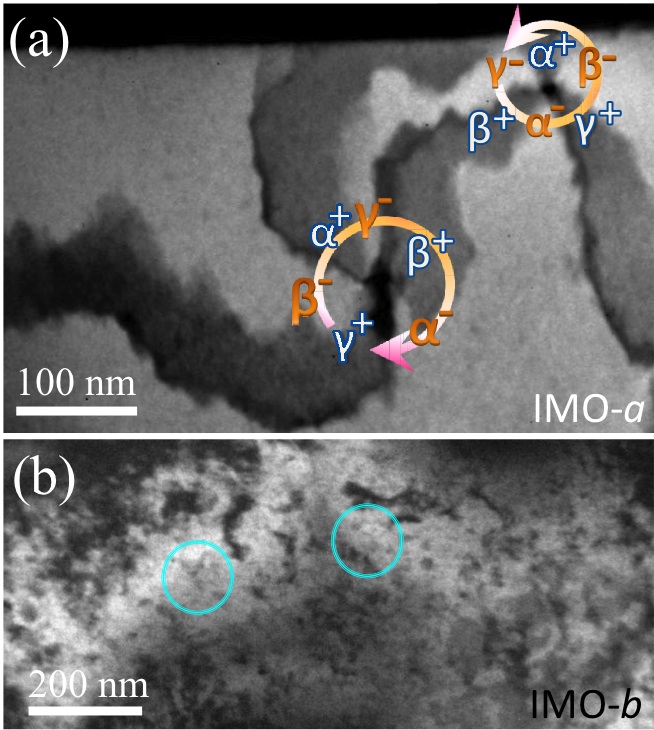}
\end{center}
\caption{\label{fig:Fig2}(Color online) (a) DF-TEM image of slow-cooled IMO-\textit{a}, taken using the \textbf{g}$^-$=($\bar{2}$22) spot, exhibits topological vortex-antivortex domains, characteristic of the FE-\textit{P}6$_3$\textit{cm} state. (b) DF-TEM image of furnace-cooled IMO-\textit{b}, taken using the \textbf{g}$^+$=(2$\bar{2}$$\bar{2}$) spot, shows numerous nanoscale speckles.}

\end{figure}                     
%======================================
%%%%%%%%%%%%%%%%%%%%%%%%%%%%%%%%%%%Fig1    
%Figure 1 two sample difference Figure~\ref{fig:Fig2}(b) shows the DF-image by exciting (2$\bar{2}$$\bar{2}$) diffraction spots where the specimens were tilted from c-axis by 16.5 degree.

%\textit{P}6$_3$\textit{cm} and \textit{P}$\bar{3}$\textit{c}1   

All four InMnO$_3$ specimens have been confirmed to show a $\sqrt{3}$$\times$$\sqrt{3}$ superstructure from the \textit{P}6$_3$/\textit{mmc} paraeletric structure in
x-ray diffraction patterns, but their electric properties exhibit surprisingly different behaviours. The \textit{P(E)} loops of IMO-\textit{a} [Fig.~\ref{fig:Fig1}(b)] indicate a clear ferroelectric hysteresis character with a remnant polarization (Pr) value of 0.73 $\mu$C/cm$^2$ while IMO-\textit{b} does not show any hint of Pr.  The drastic difference between IMO-\textit{a} and IMO-\textit{b} is consistent with the different domain morphologies revealed in dark-field TEM images. Fig.~\ref{fig:Fig2}(a) is a typical DF-TEM image of IMO-\textit{a} using the \textbf{g}$^-$=($\bar{2}$22) spot along the [101] direction based on \textit{P}6$_3$\textit{cm}, which displays cloverleaf patterns with three Mn-trimerization antiphases ($\alpha, \beta, \gamma$) coupled with opposite polarizations (+,-) \cite{Choi2010}. These cloverleaf patterns are, in fact, topological defects that are characteristic of the antiphase-ferroelectricity coupled domain configuration in hexagonal manganites \cite{Choi2010}. The alternating bright and dark contrasts result from unequal diffraction intensities associated with antiparallel polarization of the neighboring domains along the [001] direction due to the Friedel's pair breaking \cite{Gevers-F}. Depending on the sign of the vorticity, a topological defect is either a topological vortex or antivortex. These vortices and antivortices tend to be paired, and the typical size of vortex domains in IMO-\textit{a} is about 100-200 nm, which is significantly smaller than that of any \textit{h}-REMnO$_3$, showing vortex domains with the size of a few $\mu$-meters \cite{Choi2010, Chae2012}. The breaking of the inversion symmetry in IMO-\textit{a} is further confirmed in the DF-TEM image obtained using the opposite \textbf{g}$^+$=(2$\bar{2}\bar{2}$) spot [as shown in Fig. S1(a) of the Supplementary Information]. The contrasts in Fig. S1(a) are reversed from those in Fig.~\ref{fig:Fig2}(a), which unambiguously demonstrates the Friedel's pair breaking due to the non-centrosymmetrical structure of InMnO$_3$. On the other hand, the DF-TEM image of IMO-\textit{b} in Fig.~\ref{fig:Fig2}(b) shows diffusive contrasts and many nanometer-sized dark speckles indicated by sky-blue circles. It is plausible that the speckle-type pattern turns into vortex domains when a specimen was cooled slowly from the synthesis temperature. Note that the breaking of the Friedel's pairs does not take place for those speckles, indicating that those speckles contribute to the $\sqrt{3}$$\times$$\sqrt{3}$ superlattice peaks, but are not associated with the inversion symmetry breaking (see the Supplementary Information section 2).

%Note that the Pr value of 0.73 $\mu$C/cm$^2$ in IMO-\textit{a} is comparable with 6 $\mu$C/cm$^2$ in single crystalline LuMnO$_3$ [Fig.~\ref{fig:Fig1}(b)] in the sense that the Pr value in polycrystalline IMO-\textit{a} is the averaged one for all angles.
%%%%%%%%%%%%%%%%%%%%%%%%%%%%%%%%%%%Fig2

%Figure 2 sparkle feature
%fig3-HAADF-STEM     change to arrow instead of overlapping ?? zone axis                                                                                                                         
%======================================
\begin{figure}
\begin{center}
\includegraphics[width=3.0in]{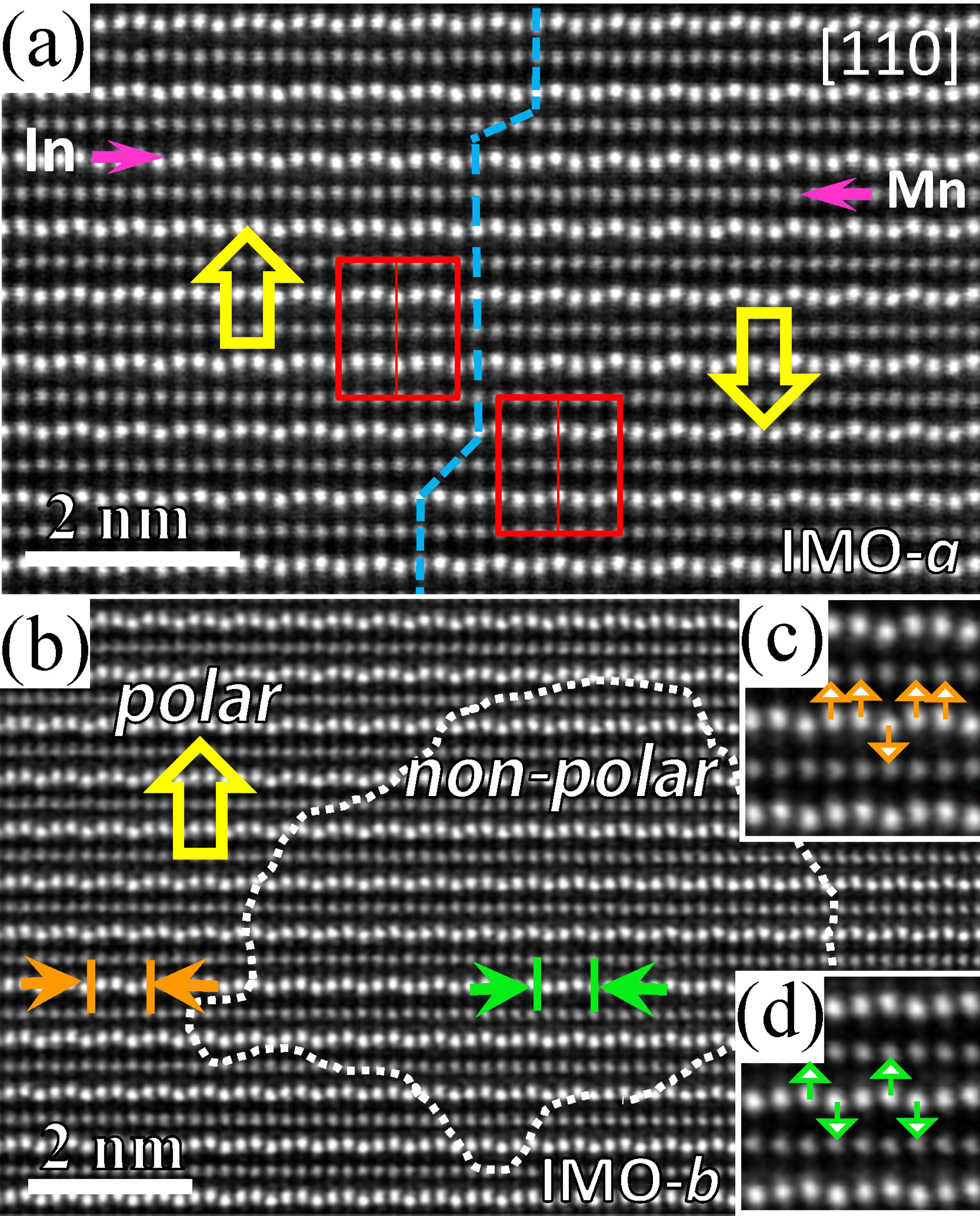}
\end{center}
\caption{\label{fig:Fig3}(Color online) HAADF-STEM image of IMO-\textit{a} (with collected angles of 80-240 mrad and the specimen thickness of 50 nm) exhibits two opposite-polarization domains with a domain wall. The unit cells are schematically shown with red rectangles, and a blue dash-line marks the domain wall with a 2\textit{a}/3 displacement (\textit{a}=5.888 {\AA}). Left and right domains show up-up-down and down-down-up In distortions, respectively. (b) HAADF-STEM image of IMO-\textit{b} shows a nanoscale non-polar domain embedded in the polar matrix. (c) and (d) display enlarged view of FE up-up-down and non-polar PUA down-no-up states. The yellow-hollow and orange/green arrows represent polarization directions and the In distortions, respectively.}
\end{figure}                     
%======================================

%The size and density of sparkles varies in a systematic manner.
%%%%%%%%%%%%%%%%%%%%%%%%%%%%%%%%%%%Fig3   

%The domain walls averaging two adjacent structural translation retain an inversion symmetry where upward and downward distortions of RE ions were cancelled \cite{Choi-2010}.
%Figure 3 
%\textit{P}6$_3$\textit{cm} and \textit{P}$\bar{3}$\textit{c}1 

In order to probe directly the atomic configuration of the ferroelectric state, we obtained HAADF-STEM images of IMO-\textit{a}, which exhibits nice vortex domains in
DF-TEM images. Fig.~\ref{fig:Fig3}(a) shows a typical HAADF-STEM image (8 nm $\times$ 5 nm) of IMO-\textit{a}, including two regions with opposite polarization
orientations (hollow arrows) with an APB-I-type domain boundary (blue dash line) \cite{Choi2010}. It is a [110] projection view with the sequence of In
(top)-Mn-In-...-In (bottom) layers, and shows bright spots corresponding to heavy In ions and weak ones corresponding to light Mn ions. Note that the assignment of In
and Mn ions has been verified by element-specific images using a STEM-EDX technique (see the Supplementary Information section 3). The In ions display clearly the
off-center shift with "up-up-down" (left) or "down-down-up" (right) distortions, but the Mn ions are well aligned without the indication of any off-center shift. The
combination of the non-centrosymmetric feature in DF-TEM images, the observation of atomic-scale up-up-down or down-down-up In distortions in HAADF-STEM images, and
the existence of the saturation polarization in \textit{P(E)} loops indicate unambiguously the presence of FE in slowly-cooled IMO-\textit{a}. On the other hand, a
HAADF-STEM image on furnace-cooled IMO-\textit{b} [Fig.~\ref{fig:Fig3}(b)] exhibits an a-few-nanometer-scale region with a non-polar "down-no-up" In configuration
(i.e., PUA, green arrows) embedded in the polar up-up-down matrix (i.e., FE, orange arrows). Fig.~\ref{fig:Fig3}(c-d) display two enlarged FE and PUA regions. The
existence of such nanoscale PUA islands, which appear corresponding to speckles in DF-TEM images [Fig.~\ref{fig:Fig2}(b)], is certainly different from the behaviour
of the thin domain walls with down-no-up distortions, which were, first, proposed as a possible configuration for the ferroelectric-trimerization domain walls in
hexagonal YMnO$_3$ \cite{Choi2010}, studied theoretically \cite{Kumagai2012}, and also observed experimentally \cite{Zhang-stem}. We emphasize that in IMO-\textit{b},
the PUA islands are very common, but the dominant phase is still the FE matrix.

%fig4-950Q + 650Q + diagram
%======================================
\begin{figure}
\begin{center}                       
\includegraphics[width=3.3in]{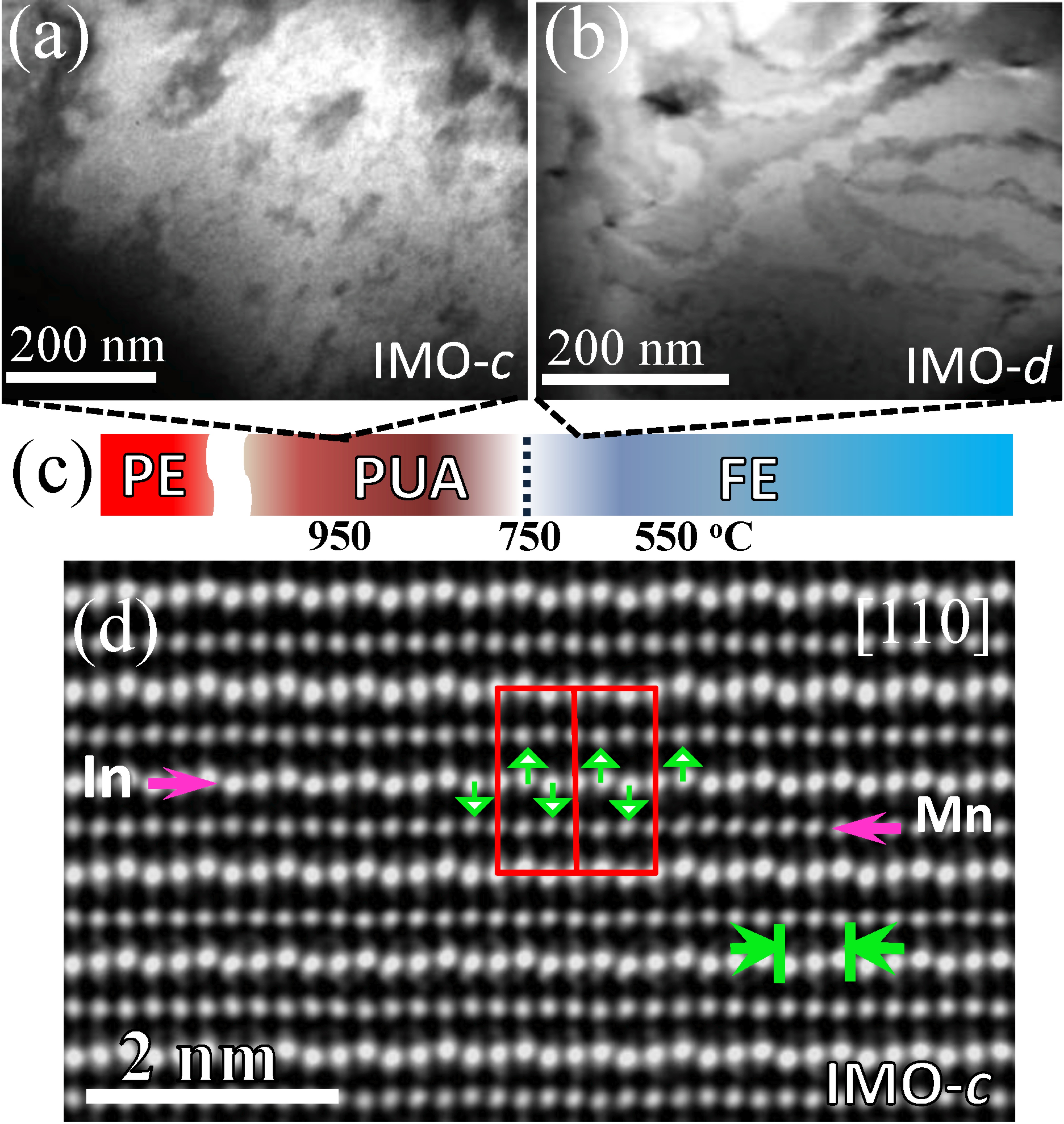}
\end{center}
\caption{\label{fig:Fig4}(Color online) (a-b) DF-TEM images of IMO-\textit{c} (quenched from 950 $^{\circ}$C) and IMO-\textit{d} (quenched from 650 $^{\circ}$C). (c) A schematic for the temperature evolution of InMnO$_3$ phases. (d) HAADF-STEM image IMO-\textit{c} shows a long-range PUA state with down-no-up In distortions. The red rectangles display the unit cells, and green arrows depict atomic In distortions.}

\end{figure}                     
%======================================
% Fig4   subgroup relation   

In order to clarify the origin of the PUA islands in IMO-\textit{b}, we performed DF-TEM and HAADF-STEM experiments on specimens from two different annealing
temperatures [950 $^{\circ}$C (IMO-\textit{c}) and 650 $^{\circ}$C (IMO-\textit{d})]. Fig.~\ref{fig:Fig4}(a) shows a DF-TEM image of IMO-\textit{c}, which exhibits no
hint of the vortex-type domains, and a HAADF-STEM image [Fig.~\ref{fig:Fig4}(d)], in fact, demonstrates a long-range PUA state\text{---}i.e., down-no-up In
distortions are everywhere in the image. On the other hand, IMO-\textit{d} exhibits small vortex-antivortex domains (50-200 nm in size) with highly curved boundaries
as shown in Fig.~\ref{fig:Fig4}(b). Further, when a specimen was quenched from 750 $^{\circ}$C, its behaviour was in-between those of IMO-\textit{c} and
IMO-\textit{d}\text{---}i.e., a mixture of very fine vortex-antivrotex domains and PUA islands. These results suggest strongly that the PUA is a stable state near 950
$^{\circ}$C, the high-T PUA can be quenched to room temperature, the transition from the high-T PUA to low-T FE is very sluggish, and the ground state is FE
\cite{Belik2009, FieBig2012}. The evolution of the structural phase in InMnO$_3$ is schematically illustrated in Fig.~\ref{fig:Fig4}(c). Note that
\textit{P}6$_3$\textit{cm} and \textit{P}$\bar{3}$\textit{c}1 are subgroups of \textit{P}6$_3$/\textit{mmc}, but \textit{P}6$_3$\textit{cm} and
\textit{P}$\bar{3}$\textit{c}1 have no subgroup relationship to each other, so it is expected that the transition between \textit{P}6$_3$\textit{cm} and
\textit{P}$\bar{3}$\textit{c}1 is 1$^{st}$-order-type.

%doping level and carrier relationship 
% for thermoelectric and topological one sentence
In summary, from comprehensive characterization of well-controlled specimens, we have clarified the long-standing dispute of the ferroelectric state of InMnO$_3$, and
identified that the ground state is a ferroelectric state with \textit{P}6$_3$\textit{cm} symmetry. In addition, the ferroelectric ground state accompanies
topological vortex domains, similar with what was observed in \textit{h}-REMnO$_3$. On the other hand, we have demonstrated the existence of an intermediate
centrosymmetric structure with \textit{P}$\bar{3}$\textit{c}1 symmetry. Furthermore, we found that the transition from the intermediate centrosymmetric to the
ferroelectric ground states is unusually sluggish. Our findings reveal the rich nature of structure-driven improper ferroelectricity.

%\acknowledgments 
The work at Rutgers was supported by the NSF under Grant No. DMR-1104484 and the National Science Council of Taiwan under project number 101-2917-I-564-077.


\begin{thebibliography}{99}   
\bibitem{Choi2010}T. Choi, Y. Horibe, H. T. Yi, Y. J. Choi, Weida Wu, and S.-W. Cheong, Nat. Mat. \textbf{9}, 253 (2010).  
\bibitem{Spaldin2004}B. B. V. Aken, T. T. M. Palstra, A. Filippetti, and N. A. Spaldin, Nat. Mat. \textbf{3}, 164 (2004).
\bibitem{Fennie2005}C. J. Fennie, and K. M. Rabe, Phys. Rev. B \textbf{72}, 100103 (2005).     
\bibitem{Katsufuji2002}T. Katsufuji \textit{et al.}, Phys. Rev. B \textbf{66}, 134434 (2002).   
\bibitem{Aken2004}B. B. VanAken, and T. T. M. Palstra, Phys. Rev. B \textbf{69}, 134113 (2004).     
\bibitem{Chae2012}S. C. Chae \textit{et al.}, Phys. Rev. Lett \textbf{108}, 167603 (2012). 
\bibitem{Abrahams2001}S. C. Abrahams, Acta. Cryst. \textbf{B57}, 485 (2001).        
\bibitem{Serrao2006}C. R. Serrao \textit{et al.}, J. Appl. Phys. \textbf{100}, 076104 (2006).  
\bibitem{Oak2011}M.-A. Oak, J.-H. Lee, H. M. Jang, J. S. Goh, H. J. Choi, and J. F. Scott, Phys. Rev. Lett \textbf{106}, 047601 (2011).
\bibitem{Belik2009}A. A. Belik, S. Kamba, M. Savinov, D. Nuzhnyy, M. Tachibana, E. Takayama-Muromachi, and V. Goian, Phys. Rev. B \textbf{79}, 054411 (2009). 
\bibitem{FieBig2012}Yu Kumagai, A. A. Belik, M. Lilienblum, N. Leo, M. Fiebig, and N. A. Spaldin, Phys. Rev. B \textbf{85}, 174422 (2012).  
\bibitem{PUA state}M. Mekata, J. Phys. Soc. Jpn. \textbf{42}, 76 (1977).
\bibitem{Tanaka-JPSJ}M. Tanaka and G. Honjo, J. Phys. Soc. Jpn. \textbf{19}, 954 (1964).
\bibitem{Gevers-F}R. Gevers, H. Blank, and S. Amelinckx, Phys. Stat. Sol. \textbf{13}, 449 (1966).
\bibitem{Kumagai2012}Y. Kumagai, and N. A. Spaldin, cond-mat. arXiv:1207.4080v1  
\bibitem{Zhang-stem}Q. H. Zhang \textit{et al.}, Phys. Rev. B \textbf{85}, 020102 (2012). 
\end{thebibliography}
\end{document}